\newcommand{\un}{~\mathrm}

\documentclass[twocolumn
,prl
,floatfix
,showpacs
,preprintnumbers
,amsmath
,amssymb
]{revtex4}

\usepackage{graphicx}
\usepackage{dcolumn}
\usepackage{bm}

\begin{document}
\title{Low self-affine exponents of fracture surfaces of glass ceramics}
\author{Laurent Ponson$^{1,2}$}\email{laurent.r.ponson@wanadoo.fr}
\author{Harold Auradou$^1$}
\author{Philippe Vi{\'e}$^3$}
\author{Jean-Pierre Hulin$^1$}
\affiliation{$^1$Laboratoire Fluides, Automatique et Syst{\`e}mes 
Thermiques, UMR 7608, Universit{\'e}s Pierre et Marie Curie-Paris 6 et Paris-Sud, 
B{\^a}timent 502, Campus Paris Sud, 91405 Orsay 
Cedex, France\\$^2$Fracture Group, Service de 
Physique et Chimie des Surfaces et Interfaces, 
DSM/DRECAM/SPCSI, B{\^a}timent 462, CEA Saclay, 
F-91191 Gif sur Yvette, France\\$^3$Laboratoire des mat{\'e}riaux et structures du g{\'e}nie civil, UMR 113, all{\'e}e K{\'e}pler, Marne la Vall{\'e}e Cedex.}

\begin{abstract}
The geometry of post mortem rough fracture surfaces of porous glass ceramics made of sintered glass beads is shown experimentally 
to be self-affine with an exponent  $\zeta=0.40\, \pm\, 0.04$ remarkably lower than the 'universal' value $\zeta=0.8$ frequently 
measured  for many materials.   This low value of $\zeta$ is similar to that found for sandstone samples of similar micro structure 
and is also practically independent on the porosity $\phi$ in the range investigated ($3\un{\%} \le \phi \le 26\un{\%}$) as well as  on the 
bead diameter $d$ and of the crack growth velocity. In contrast, the roughness amplitude normalized by  $d$  increases linearly 
with $\phi$ while it is still independent, within experimental error, of $d$ and of the crack propagation velocity. An interpretation 
of this variation is suggested in terms of a transition from transgranular to intergranular fracture propagation with no influence, 
however, on the exponent $\zeta$.
\end{abstract}

\date{\today}
\maketitle

Since the pioneering work of Mandelbrot \cite{Mandelbrot}, statistical mechanics has been used to describe fracture processes. 
A special attention has been brought  to the statistical properties of the roughness of fracture surfaces which often display a 
scaling behavior referred to as {\it self-affinity}. This means that  they are statistically invariant under the 
transformation : $\bf{r} \rightarrow \lambda \bf{r}$ and $h \rightarrow \lambda^\zeta h$; ($\bf{r}$ characterizes a location 
in the mean plane of the fracture, $h$ is the surface elevation and $\zeta$ is the self-affine exponent~\cite{Feder}). Due to 
this invariance, the height-height correlation function $\Delta h(\Delta r)=<(h({\bf r}+{\bf \Delta r}) - h({\bf r}))^2>^{1/2}_{\bf r}$ scales like :
\begin{equation}
\frac{\Delta h}{\Delta h(d)} = \left(\frac{\Delta r}{d}\right)^\zeta,
\label{eq:eq1}
\end{equation}
 $\bf{\Delta r}$ is a vector of constant orientation and of  modulus $\Delta r$; $d$ will  be taken equal to the characteristic length scale 
of the microstructure of the material (the mean grain size for instance). As a consequence, the amplitude of the roughness may be characterized 
by the value of $\Delta h(d)$ (amplitudes corresponding to other $\Delta r$ values can then be obtained from Eq.~(\ref{eq:eq1})). 
 
Experimental studies of various fractured $3D$ materials 
provided similar values of roughness exponents $\zeta \simeq 0.8$ (see Ref. \cite{Bouchaud} for a review). These results supported 
the idea of a universal roughness exponent $\zeta$ independent on the material. However, sandstone fracture surfaces were found to be self-affine 
with a lower exponent close to $0.5$ \cite{Boffa}. Explaining these discrepancies  is important to understand  the physical origin of their {\it self-affine} 
geometry. The objective of the present paper is to analyze these effects on an artificial material comparable to sandstone but with a microstructure that can be 
varied  in a controlled manner.

Glass ceramics made of sintered glass beads have been selected for that purpose because, like sandstone, they are made of cemented grains. However, 
in contrast with natural rocks, both the characteristic size and the cohesion of the grains can be adjusted by modifying respectively the bead diameter 
and the sample porosity $\phi$.  The exponent $\zeta$ for these materials will be shown to be much lower  than the "universal value" $0.8$; in addition, 
it depends surprisingly  little on the microstructure of the material, although the amplitude of the roughness varies significantly with the porosity. 
The influence of velocity on roughness reported by some authors \cite{Daguier}will also be evaluated by comparing two types of mode I 
fracturing processes with different crack propagation velocities.

The samples are prepared by heating a mold filled with a random packing of beads at $700^\circ C$ during $20$ to $200$ minutes. Depending on 
the duration of the annealing, the degree of sintering varies and porosities $\phi$ ranging from $3\un{\%}$ to $26\un{\%}$ are obtained. Either of two sets 
of beads with diameters in the range $104-128~\mu\mathrm{m}$ and $50-65~\mu\mathrm{m}$ are used. In the following, the characteristic microscopic 
length scale $d$ introduced in Eq. (\ref{eq:eq1}) is taken equal to the mean bead diameter. The open porosity is measured by saturating the sample with 
water and the profile of the total porosity along the sample is measured by $\gamma$-ray absorption. Porosity variations along the samples 
are of the order of $1\un{\%}$ so that $\phi$ is considered as constant and equal to this mean value within $\pm 1\%$. The process provides sintered glass 
cylinders with a $130\un{mm}$ height and a $40\un{mm}$ radius from which the samples  used in the fracture tests are cut out. 

Two mode $I$ fracture tests have been performed. Fast fracture propagations are obtained by means of modified Brazilian fracture tests using toroidal 
samples (inside and outside radii : $10\un{mm}$ and $40\un{mm}$ respectively; height parallel to the axis :  $30\un{mm}$). A uniaxial 
radial compression is applied to the torus and two symmetrical dynamical cracks propagate from the central hole toward the outer zones where the 
compressive forces are applied.

Quasistatic fracture propagations are realized on samples  with a  modified Tapered Double Cantilever Beam (TDCB) geometry : Their width and 
length are respectively $20\un{mm}$ (perpendicular to the crack propagation) and $60\un{mm}$ (parallel to it). The fracture is 
initiated from a straight notch  (thickness $1\un{mm}$)  by applying on both sides a uniaxial tension with a constant opening rate : The 
tapered shape of these specimens allows us to obtain a stable mode $I$ crack growth (see Ref. \cite{Liebowitz}). The crack propagation 
velocity $V_f$, determined from the variations  of the electrical resistance of a thin gold layer deposited on the side of the 
sample, remains close to $3\un{mm.s^{-1}}$.

\begin{figure}[!h]
\includegraphics[width=0.75\columnwidth]{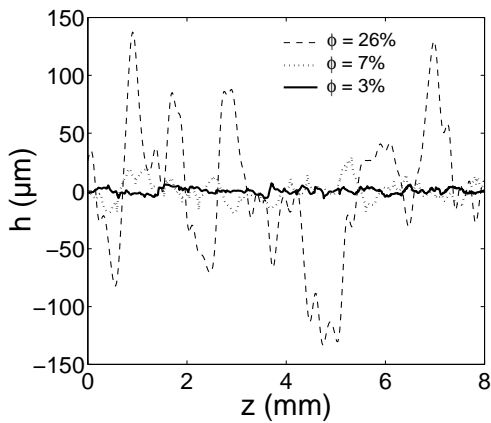}
\centering
\caption{Height profiles extracted from fracture surfaces of three sintered glasses with three different porosities but with beads with 
the same diameter $104-128~\mu\mathrm{m}$.}
\label{profiles}
\end{figure}
 Fig. \ref{profiles} displays fracture surface profiles of three samples of different porosities. The roughness amplitude increases by almost 
two orders of magnitude from $\simeq 1~\mu\mathrm{m}$ for  the densest ($\phi = 3\un{\%}$) to $\simeq 100~\mu\mathrm{m}$ for the most porous 
sample ($\phi = 26\%$) : Different profilometers are therefore needed to scan all the samples. For surfaces with a very low spatial 
roughness (and therefore porosity) we use an interferometric optical profilometer ({\texttrademark}Talysurf CCI 6000) with a vertical 
resolution better than $0.1\un{nm}$  and a lateral resolution $\simeq 1~\mu\mathrm{m}$. The local slope of the surface must always 
be small to apply this technique so that, for $\phi > 7\un{\%}$,
 a mechanical stylus profilometer ({\texttrademark}Talysurf Intra) with a $\simeq 10\un{nm}$ vertical and a $\simeq 2~\mu\mathrm{m}$ lateral resolutions had to be used. For lower porosities, both profilometers are usable
which allowed to check the consistency of the two measurements. For both profilometers the maps contain 
 $1024 \times 1024$ points and the fields of view are respectively $3\times 3$ and
  $6\times 6\un{mm}$.
In the second technique, the stylus is always in contact with the surface and, for porosities higher than $18\%$,  it often gets  jammed into 
the deepest asperities of the surface. One uses then a point by point mechanical profilometer : A sensor tip is lowered down until it touches 
the surface in order to measure its height; the tip is then  raised by $200~\mu\mathrm{m}$ before getting moved laterally by $25~\mu\mathrm{m}$ to 
the next measurement point.  The vertical and the lateral resolutions are respectively $\simeq 3~\mu\mathrm{m}$ and  $\simeq 10~\mu\mathrm{m}$ and 
the typical field of view is $6\times 6\un{mm}$. The consistency of the measurements was verified by comparing profiles provided by the two 
mechanical systems for $\phi = 18\un{\%}$.

These measurements provide  surface elevation profiles as shown in  Fig. \ref{profiles}.
In the following, we shall only analyze profiles  parallel to the $z$-axis ({\it i.e.} to the crack front) and located far enough from the 
initiation region so that the roughness properties are  statistically stationary (this latter point is discussed in more detail below.) 

\begin{figure}[!h]
\includegraphics[width=0.75\columnwidth]{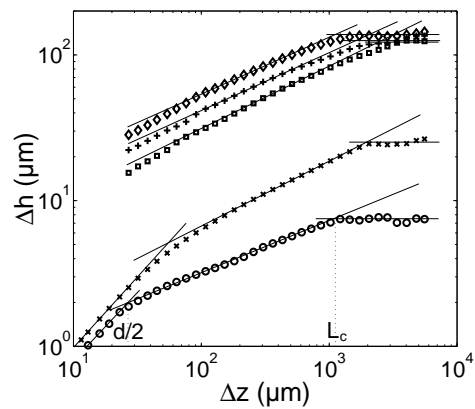}
\centering
\caption{Log-log representation of $\Delta h$ (averaged over profiles at different $x$ values) as a function of $\Delta z$ for surface 
roughness profiles of several sintered glass samples. 
- Samples fractured using the TDCB test with  porosities $\phi=7\un{\%}$ ($\times$)  , $15\un{\%}$ ($\square$) and $26\un{\%}$ ($+$)  - Samples 
fractured using the Brazilian test  with porosities  $\phi=3\un{\%}$ ($\circ$) and $25\un{\%}$ ($\diamond$)
- Range of bead diameters used to realize samples : $104-128~\mu\mathrm{m}$
- Straight lines : Linear fits with slope $\zeta$ (Tab. \ref{tab:tab1}).}
\label{fig:fig2}
\end{figure}
Quantitatively, these profiles will be characterized by their  $1$D height-height correlation 
function, $\Delta h(\Delta z) = <(h(z+\Delta z) - h(z))^2>^{1/2}_z$ :  For surfaces with a  {\it self-affine} roughness, the power 
law relationship (\ref{eq:eq1}) should be verified. 
This relation is tested in Fig. \ref{fig:fig2} which displays in a log-log scale the variations of $\Delta h(\Delta z)$ as a function of $\Delta z$ for 
several samples : These have different porosities in the range : $3\un{\%} \le \phi \le 26\un{\%}$ and have been fractured both in a quasistatic mode and by 
the modified Brazilian test. The curve corresponds to an average of $\Delta h(\Delta z)$  over profiles corresponding to different distances $x$ to 
the initiation and  located in the region where  the statistics of the roughness is stationary. 
 
Let us examine, for instance, the lower curve corresponding to a $3\un{\%}$ porosity sample fractured using the modified Brazilian test procedure. 
The variation can clearly not be fitted by a single power law over the full range of $\Delta z$ values investigated : Three domains of variation are 
visible and correspond to exponents respectively equal to $1$, $0.36$ and $0$ (straight lines). The last value indicates that the surface appears as 
a plane at large scales and the first one corresponds to an Euclidian geometry of individual grains. The surface profile is therefore self-affine 
(here with an exponent $\zeta = 0.36$) only in the intermediate domain between two limiting length scales : In the log-log plot of Fig.~\ref{fig:fig2}, 
they correspond to the intersections between the straight lines fitted in the different  domains. The lower boundary is of the order of the bead 
radius $d/2$ and the upper one will be referred to as $L_c$. 

This result is generalized by comparing the different curves in Fig. \ref{fig:fig2} corresponding to samples of different porosities and fractured both 
in the quasistatic and fast propagation modes. 
All the curves have the same global shape and,  in log-log coordinates, the slopes are nearly  the same in the intermediate domain : This shows that 
the roughness exponent is very similar in all cases while the vertical shift between the curves reflects different roughness amplitudes.

\begin{table}[!h]
\begin{center}
\begin{tabular}{|l|c|c|c|c|c|c|}
 \hline
  & $\phi=3\%$ & $\phi=7\%$ & $\phi=15\%$ & $\phi=18\%$ & $\phi=25\%$ & $\phi=26\%$ 
   \\ \hline \hline  & B.D.$1$ & B.D.$1$ & B.D.$1$ & B.D.$2$ & B.D.$1$ & B.D.$1$
   \\ \hline & $Dyn.$ & $Q.S.$ & $Q.S.$ & $Dyn.$ & $Dyn.$ & $Q.S.$        
    \\ \hline Met. & $1-2$ & $2$ & $3$ & $2-3$ & $3$ & $3$        
  \\ \hline \hline $\zeta$ & $0.36$ & $0.43$ & $0.43$ & $0.40$ & $0.40$ & $0.39$
   \\ \hline $\zeta_{FT}$ & $0.38$ & $0.44$ & $0.38$ & $0.37$ & $0.39$ & $0.39$
   \\ \hline $\Delta h(d)$   & 3.8 & 8.8 & 32 & 16 & 56 & 44
   \\ \hline $L_c$   & 1.1 & 1.9 & 2.4 & 0.8 & 1.1 & 1.6
     \\ \hline
\end{tabular}
\caption{Physical and statistical characteristics of the samples - ($\phi$)  sample porosity; glass beads diameter range (B.D.$1$)  $104-128~\mu\mathrm{m}$, 
(B.D.$2$)  $50-65~\mu\mathrm{m}$; crack propagation mode   ($Dyn.$)  dynamic, ($Q.S.$) quasi static; surface measurement technique ($1$)  interferometric, ($2$)  
stylus profilometer ($3$) point by point - Statistical characteristic parameters of 1D profiles normal to crack propagation ($\zeta\,,\,\zeta_{FT}$)  
self-affine exponent values obtained respectively from the variation of $\Delta h$ with $\Delta z$ and from the Fourier power spectrum; ($\Delta h(d)$)  
roughness amplitude in $\mu\mathrm{m}$;  ($L_c$) upper boundary of self-affine domain in $\mathrm{mm}$.}
\label{tab:tab1}
\end{center}
\end{table}

The numerical values of the parameters characterizing all these curves are listed in 
Table~\ref{tab:tab1} for all samples  investigated in the present work. Table~\ref{tab:tab1} confirms  that $\zeta$ has a very similar 
value $\zeta \,=\,0.40 \pm 0.04$ for all samples independent of the bead size, of the porosity and of the crack propagation velocity: This 
common value is much lower that the value $0.8$ reported for many materials and closer to the value $0.5$ obtained for sandstone.  This result 
is robust with respect to the method used to determine $\zeta$ as shown by the comparison with the values $\zeta_{FT}$ in Table~\ref{tab:tab1} 
obtained from the analysis of power spectra~\cite{Feder}. The upper boundary $L_c$ of the self-affine domain is of the order of $1-2~\mathrm{mm}$ 
i.e. about a tenth of the sample width and seems to decrease for faster fracture propagations. A similar limitation of the self-affine  range of 
length scales by  finite size effects has recently been reported~\cite{Mourot3} in experiments on mortar samples of different widths. 

\begin{figure}[!h]
\includegraphics[width=0.75\columnwidth]{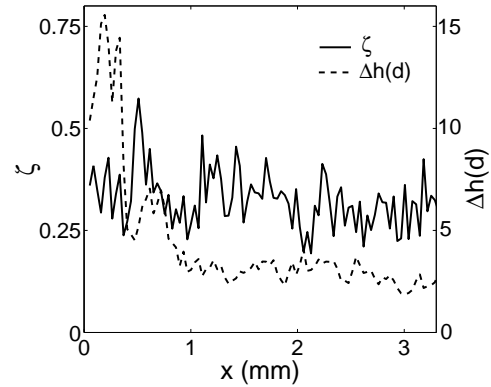}
\centering
\caption{Variations of the roughness characteristic exponent $\zeta$ and amplitude $\Delta h(d)$ with the distance $x$ to the initiation notch 
for the same sample ($\phi=3\%$) as in Fig.\ref{fig:fig2}. }
\label{fig:fig3}
\end{figure}

In the above interpretations, it is assumed that, after a large enough propagation distance from the initiation, the statistical properties of the 
roughness become constant.  In order to verify this assumption, the transient regime already reported  by other authors~\cite{Schmittbuhl2, Morel} is  
studied in Fig.~\ref{fig:fig3}. In this figure, data points correspond to single profiles parallel to $z$ at a given distance $x$ from the initial 
notch. The roughness of each profiles is characterized by its amplitude $\Delta h(d,x)$ and by the exponent  $\zeta(x)$. For $x<1\ mm$, a 
transient regime in which  $\Delta h(d,x)$ decreases with the distance $x$ while $\zeta(x)$ remains constant is indeed observed. At larger distances,  $\zeta(x)$ and $\Delta h(d,x)$ merely fluctuate around an average value. For the 
other samples investigated, the distance has the same order of magnitude. These results allow to restrict (as above) the statistical analysis  
to distances $>1\un{mm}$ where the roughness is statistically stationary : This justifies our use in Fig.~\ref{fig:fig2} of curves 
obtained from averages over several profiles corresponding to different $x$ values. 
\begin{figure}[!h]
\includegraphics[width=0.75\columnwidth]{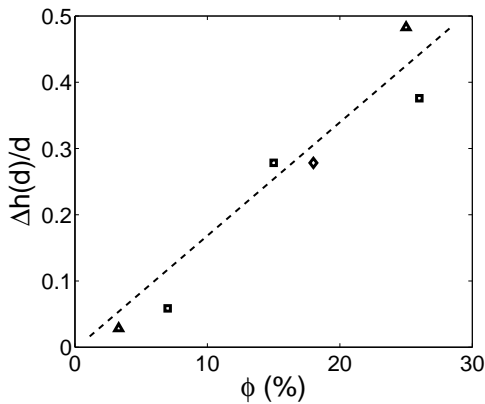}
\centering
\caption{Variation of $\Delta h (d)/d$ as a function of porosity $\phi$. - Range of bead diameters used to realize 
samples -$104-128~\mu\mathrm{m}$ ($\triangle$) : Dynamic fracture ($\square$) : Quasistatic fracture -  $50-65~\mu\mathrm{m}$ ($\lozenge$) dynamic 
fracture. The slope of the dashed line is $1.7$.}
\label{fig:fig4}
\end{figure}

The curves of Fig. \ref{fig:fig3} have confirmed the extreme robustness of the exponent $\zeta$ which remains constant even in the transient regime.
In contrast, the roughness amplitude $\Delta h(d)$ appears to be more sensitive, not only to  $x$, but also to the characteristic parameters of the sample 
like its porosity as seen in Fig. \ref{profiles}. 
This variation is analyzed quantitatively in Fig. \ref{fig:fig4}, in which  the normalized variable $\Delta h(d)/d$ is plotted as a function of the 
porosity $\phi$. For low porosities, $\Delta h(d)/d$ 
decreases to zero while it is close to $0.5$ for the most porous sample and a good collapse of all data on a linear variation is observed. This suggests that the velocity $V_f$ of the crack propagation has a weak influence on the amplitude $\Delta h(d)$ of the roughness and 
that $\Delta h(d)$ is roughly proportional both to the bead size $d$ and to $\phi$. 

These results and, particularly, the increase of $\Delta h(d)/d$ with $\phi$ may be related to phenomena at the scale of a bead diameter. For high porosity 
samples, cracks propagate by breaking cemented necks binding two beads : The difference in height between neighboring beads is then of the order of their 
radius and $\Delta h(d)/d \simeq 0.5$. For low porosity samples, neighboring beads are more strongly welded to each other and cracks propagate through 
the beads : The deflections of the surface are weaker compared to the bead radius and $\Delta h(d)/d$ is lower. This is consistent with the increase 
of $\Delta h(d)/d$ with $\phi$ in Fig. \ref{fig:fig4}. These differences in the propagation of the cracks are confirmed by scanning electron 
microscope (SEM) images of fractured samples : These display a transition from transgranular to intergranular propagation as the porosity 
increases. It is remarkable that this transition has no influence on the characteristic exponent $\zeta$.

To conclude, the fracture surfaces of sintered glass samples analyzed in the present work all display  a self-affine domain of length scales characterized by 
an exponent $\zeta=0.40\pm 0.04$ : This value of $\zeta$ has been found to be independent of the porosity $\phi$, of 
the fracture propagation velocity $V_f$ and of the diameter $d$  of the beads. $\zeta$  is significantly lower than for many materials ($\zeta \simeq 0.8$) 
but close to the value found for sandstone ($\zeta \simeq 0.5$) which has a similar structure. This nearly constant value of $\zeta$ contrasts with the 
linear increase with $\phi$ of the normalized roughness amplitude $\Delta h(d)/d$ (although it is independent both of $V_f$ and $d$).
 This  increase of the amplitude likely reflects a progressive transition  from transgranular to intergranular fracture propagation also demonstrated by SEM 
observations : A surprising feature is the fact  that this transition does not influence $\zeta$.
Another important issue is the anisotropy of the surface roughness : Preliminary measurements of profiles parallel to the fracture propagation have given 
a value $\zeta \simeq 0.5$ larger than that reported above for profiles parallel to the front.  Similar differences have been reported \cite{Ponson5,Ponson4}  
on other materials using a 2D analysis  based on  the computation of a 2D correlation function : Applying this latter technique to the present samples might  
help confirm the presence and amplitude of the anisotropy.
These results may be compared to the theoretical  predictions of brittle fracture models describing crack growth as the propagation of an elastic line in random 
heterogeneous media~\cite{JPBouchaud, Ramanathan, Bonamy2}, leading to $\zeta=0.39$ in the 2D case \cite{Schmittbuhl4, Rosso2} and the 3D case \cite{Bonamy2}. 
Ductile fracture models such as those developed in \cite{Hansen} and leading to higher roughness exponent values fail to reproduce the observations reported here 
because damage processes are not expected during the failure of glass ceramics. Further studies are currently in progress to confirm that these models are 
applicable to such materials as glass ceramics or sandstone displaying brittle fracture and whether the higher value $\zeta=0.8$ measured for many others may be 
associated to ductile fracture.

\begin{acknowledgments}
Fracture surface scans were realized  in cooperation with C. Buisson (Taylor-Hobson France). We thank S. Roux and Saint-Gobain Recherche for providing us 
with SEM pictures.
We are grateful to A. Azouni, D. Bonamy, E. Bouchaud, K. M{\aa}l{\o}y, S. Morel, S. Roux and N. Shahidzadeh-Bonn for many enlightening discussions. 
\end{acknowledgments}

\end{document}